\documentclass[conference,a4paper]{IEEEtran}
\IEEEoverridecommandlockouts
\usepackage[style=ieee,
  isbn=false,
  doi=false,
  sorting=none,
  url=true,
  defernumbers=true,
  bibencoding=utf8,
  backend=biber,
  mincitenames=1, maxcitenames=3,
  minbibnames=1, maxbibnames=24]{biblatex}
\usepackage{amsmath,amssymb,amsfonts}
\usepackage{soul}
\usepackage{multicol}
\usepackage{algorithm}
\usepackage{algpseudocode}
\usepackage{booktabs}
\usepackage{datetime}
\usepackage{graphicx}
\usepackage{textcomp}
\usepackage[hidelinks]{hyperref}
\usepackage{xcolor}
\usepackage{tabularx}
\usepackage{todonotes}
\usepackage{standalone}
\usepackage{balance}
\usepackage[inkscapelatex=false]{svg}
\usepackage{url}
\usepackage[nolist]{acronym}
\def\BibTeX{{\rm B\kern-.05em{\sc i\kern-.025em b}\kern-.08em
    T\kern-.1667em\lower.7ex\hbox{E}\kern-.125emX}}
\usepackage{siunitx}

\newcommand{\byte}{B}

\renewcommand\IEEEkeywordsname{Keywords}

\algdef{SE}[CALLBACK]{Callback}{EndCallback}
   {\algorithmiccallback}
   {\algorithmicend\ \algorithmiccallback}
\algnewcommand{\algorithmiccallback}{\textbf{callback} }

\algdef{SE}[ISR]{ISR}{EndISR}
   {\algorithmicisr}
   {\algorithmicend\ \algorithmicisr}
\algnewcommand{\algorithmicisr}{\textbf{isr} }

\newcommand{\wifi}{Wi\nobreakdash-Fi}

\makeatletter
\let\old@ps@headings\ps@headings
\let\old@ps@IEEEtitlepagestyle\ps@IEEEtitlepagestyle
\def\confheader#1{%
\def\ps@headings{%
\old@ps@headings%
\def\@oddhead{\strut\hfill#1\hfill\strut}%
\def\@evenhead{\strut\hfill#1\hfill\strut}%
}%
\def\ps@IEEEtitlepagestyle{%
\old@ps@IEEEtitlepagestyle%
\def\@oddhead{\strut\hfill#1\hfill\strut}%
\def\@evenhead{\strut\hfill#1\hfill\strut}%
}%
\ps@headings%
}
\makeatother
\confheader{%
2023 13th International Conference on Indoor Positioning and Indoor Navigation (IPIN)}

\newcommand{\added}[1]{\textcolor{black}{#1}}

\begin{document}

\hyphenpenalty   10000
\exhyphenpenalty 10000
\widowpenalty    10000
\clubpenalty     10000

\title{UJI Probes: Dataset of \wifi{} Probe Requests
\thanks{The authors gratefully acknowledge funding from European Union’s Horizon 2020 Research and Innovation programme under the Marie Sk\l{}odowska~Curie grant agreement No. $813278$ (A-WEAR: a~network for dynamic wearable applications with privacy constraints, \protect{\url{http://www.a-wear.eu/})} and No. $101023072$ (ORIENTATE: Low-cost Reliable Indoor Positioning in Smart Factories, \url{http://orientate.dsi.uminho.pt/}). This work does not represent the opinion of the European Union, and the European Union is not responsible for any use that might be made of its content.}}

\author{\IEEEauthorblockN{Tomáš Bravenec\IEEEauthorrefmark{1}\IEEEauthorrefmark{2},
Joaquín Torres-Sospedra\IEEEauthorrefmark{3}, Michael Gould\IEEEauthorrefmark{1} and Tomas Fryza\IEEEauthorrefmark{2}}
\IEEEauthorrefmark{1}\textit{Institute of New Imaging Technologies, Universitat Jaume I}, Castellón, Spain \\
\IEEEauthorrefmark{2}\textit{Department of Radio Electronics, Brno University of Technology}, Brno, Czech Republic \\
\IEEEauthorrefmark{3}\textit{Algoritmi Research Centre, University of Minho}, Guimarães, Portugal\\ \texttt{bravenec@uji.es} -- \texttt{jtorres@algoritmi.uminho.pt} -- \texttt{gould@uji.es} -- \texttt{fryza@vut.cz}}

\maketitle

\begin{abstract}
This paper focuses on the creation of a~new, publicly available \wifi{} probe request dataset. \added{Probe requests belong to the family of management frames used by the 802.11 (Wi-Fi) protocol.} As the situation changes year by year, and technology improves probe request studies are necessary to be done on up-to-date data. We provide a~month-long probe request capture in an office environment, including work days, weekends, and holidays consisting of over 1 400 000 probe requests. We provide a~description of all the important aspects of the dataset. Apart from the raw packet capture we also provide a~\ac{rem} of the office to ensure the users of the dataset have all the possible information about the environment. To protect privacy, user information in the dataset is anonymized. This anonymization is done in a~way that protects the privacy of users while preserving the ability to analyze the dataset to almost the same level as raw data. Furthermore, we showcase several possible use cases for the dataset, like presence detection, temporal \ac{rssi} stability, and privacy protection evaluation.
\end{abstract}

\acresetall


\begin{acronym}[A-WEAR]
    \acro{5g}[5G]{Fifth-Generation}
    \acro{aoa}[AoA]{Angle of Arrival}
    \acro{aod}[AoD]{Angle of Departure}
    \acro{aoi}[AoI]{Area of Interest}
    \acro{ann}[ANN]{Artificial Neural Network}
    \acro{ap}[AP]{Access Point}
    \acro{awear}[A-WEAR]{A network for dynamic WEarable Applications with pRivacy constraints}
    \acro{awknn}[aw$k$NN]{Adaptive Weighted $k$-Nearest Neighbors}
    \acro{ble}[BLE]{Bluetooth Low Energy}
    \acro{cdf}[CDF]{Cumulative Distribution Function}
    \acro{cnn}[CNN]{Convolutional Neural Network}
    \acro{cpu}[CPU]{Central Processing unit}
    \acro{csi}[CSI]{Channel State Information}
    \acro{csv}[CSV]{Comma Separated-Values}
    \acro{cv}[CV]{Computer Vision}
    \acro{d2d}[D2D]{Device-to-Device}
    \acro{dsss}[DSSS]{Direct Sequence Spread Spectrum}
    \acro{dt}[DT]{Digital Twins}
    \acro{dwfwknn}[dwfw$k$NN]{Distance \& Feature Weighted $k$-Nearest Neighbors}
    \acro{ejd}[EJD]{European Joint Doctorate}
    \acro{esr}[ESR]{Early Stage Researcher}
    \acro{fhss}[FHSS]{Frequency Hopping Spread Spectrum}
    \acro{gdpr}[GDPR]{General Data Protection Regulation}
    \acro{gnss}[GNSS]{Global Navigation Satellite Systems}
    \acro{gpr}[GPR]{Gaussian Process Regression}
    \acro{gpu}[GPU]{Graphical Processing Unit}
    \acro{hid}[HID]{Human Interface Device}
    \acro{hmm}[HMM]{Hidden Markov Models}
    \acro{ips}[IPS]{Indoor Positioning System}
    \acro{iot}[IoT]{Internet of Things}
    \acro{ir}[IR]{Infrared}
    \acro{isr}[ISR]{Interrupt Service Routine}
    \acro{itn}[ITN]{Innovative Training Network}
    \acro{idw}[IDW]{Inverse Distance Weight}
    \acro{knn}[$k$NN]{$k$-Nearest Neighbors}
    \acro{lid}[LID]{Linearly Interpolated Data}
    \acro{los}[LoS]{Line of Sight}
    \acro{md}[MD]{Measured Data}
    \acro{ml}[ML]{Machine Learning}
    \acro{mac}[MAC]{Media Access Control}
    \acro{nato}[NATO]{North Atlantic Treaty Organization}
    \acro{nlos}[NLoS]{Non Line of Sight}
    \acro{ntp}[NTP]{Network Time Protocol}
    \acro{ofdm}[OFDM]{Orthogonal Frequency Division Multiplexing}
    \acro{oui}[OUI]{Organizationally Unique Identifier}
    \acro{pan}[PAN]{Personal Area Network}
    \acro{pir}[PIR]{Passive Infrared}
    \acro{pnl}[PNL]{Preferred Network List}
    \acro{poi}[PoI]{Point of Interest}
    \acro{ptp}[PtP]{Point-to-Point}
    \acro{ram}[RAM]{Random Access Memory}
    \acro{rbf}[RBF]{Radial Basis Function}
    \acro{relu}[ReLU]{Rectified Linear Unit}
    \acro{rem}[RM]{Radio Map}
    \acro{rfid}[RFID]{Radio Frequency Identification}
    \acro{rnn}[RNN]{Recurrent Neural Network}
    \acro{ro}[RO]{Research Objective}
    \acro{rp}[RP]{Reference Point}
    \acro{rq}[RQ]{Rational Quadratic}
    \acro{rssi}[RSSI]{Received Signal Strength Indicator}
    \acro{rtc}[RTC]{Real Time Clock}
    \acro{sawknn}[saw$k$NN]{Self-Adaptive Weighted $k$-Nearest Neighbors}
    \acro{se}[SE]{Squared Exponential}
    \acro{ssid}[SSID]{Service Set Identifier}
    \acro{stiknn}[sti-$k$NN]{Signal Tendency Index -- Weighted $k$-Nearest Neighbors}
    \acro{svm}[SVM]{Support Vector Machines}
    \acro{toa}[ToA]{Time of Arrival}
    \acro{tof}[ToF]{Time of Flight}
    \acro{tdoa}[TDoA]{Time Difference of Arrival}
    \acro{ue}[UE]{User Equipment}
    \acro{uuid}[UUID]{Universal Unique Identifier}
    \acro{uuide}[UUID-E]{Universally Unique IDentifier-Enrollee}
    \acro{uwb}[UWB]{Ultra-Wideband}
    \acro{wknn}[w$k$NN]{Weighted $k$-Nearest Neighbors}
    \acro{wlan}[WLAN]{Wireless Local Area Network}
    \acro{wps}[WPS]{Wi-Fi Protected Setup}
    \acro{wwan}[WWAN]{Wireless Wide Area Network}
    \acro{xr}[XR]{eXtended Reality}
\end{acronym}

\section{Introduction}

\added{The growing interest in indoor positioning and indoor navigation also raises questions related to privacy. Unlike \ac{gnss}, where the whole world uses the same coordinate systems and \acp{ue} can calculate their own location from the received information, indoor location in most cases requires cooperation between the \ac{ue} and the positioning infrastructure. This creates a~window of opportunity, through which the privacy of users can be breached. This is especially true when it comes to \wifi{} since the management frames used for probing the environment for nearby \acp{ap} are not encrypted.}

\added{These m}anagement frames of \wifi{} have been the center of attention of researchers for years now. Back before the introduction of \ac{mac}, \textcite{musa2012tracking} used the real \ac{mac} addresses in probe requests for mobility tracking. Then the Sapienza~probe request dataset~\cite{barbera2013signals, barbera2013crawdad} was published including the analysis of social relationships from the probe requests. Before the implementation of \ac{mac} address randomization~\textcite{cunche2014linking, cheng2013characterizing} showed the vulnerability of probe requests to tracking techniques. Then in 2014, the \ac{mac} address randomization was introduced by Apple in the iOS version 8~\cite{ios8_hutchinson_2014}.

Since Apple introduced the randomization of \ac{mac} addresses, the research community explored the weaknesses of this privacy-related measure. \textcite{martin2016decomposition} explored the ways of revealing the globally unique \ac{mac} addresses of devices employing \ac{mac} address randomization. Just a~year later \textcite{freudiger2015talkative} worked on reverse engineering \ac{mac} address randomization. In 2016, \textcite{di2016mindYourProbes} analyzed probe requests captured in Italy during political events, trying to infer the origin of people participating in the events. Their results did indeed match the officially published voting reports. There are also other possible ways to analyze probe requests: \textcite{matte2016defeating} exploited the time difference between subsequent probe requests to identify different devices. In 2021 \textcite{fenske2021three} published a~deep look into user privacy protection, which was created as a~follow up to a~similar study published few years before it by \textcite{martin2017study}.

\subsection{Probe Requests}

Probe requests belong to the management frames family of the IEEE 802.11 standard~\cite{ieee802.11aq_standard}. These frames are specific in that they are sent without any encryption and any device with monitoring mode available in their wireless interface driver can capture and read these frames. The primary purpose of probe requests is the detection of nearby \wifi{} networks. Primarily its purpose was to find a~known \ac{ap} to connect to, but a~secondary purpose came with the popularity of \wifi{}. That is for acquiring a~rough location estimate, by matching the \ac{ssid} of nearby \ac{ap} to either public database of \ac{ap}s~\cite{wiggle} or private database of the system developer~\cite{google_location_api, google_optout_wifi}.

The probe requests consist of a~header and an information element. The header is not probe request specific and contains the frame control number (4 for probe requests), \ac{mac} addresses of the destination and source devices (with destination address being most often broadcast address - \textit{ff:ff:ff:ff:ff:ff}) and sequence number of the frame. When the broadcast address is used, the packet is targeted to every device in the network, or in the case of wireless communications to every device in the proximity. This is important so every \wifi{} \ac{ap} receives the probe request. There can be a~lot of information included in the probe requests in the information element section. It can contain \ac{ssid}, in cases the device is searching for a~specific network, otherwise, this field can stay empty. The rest of the fields contain information about the supported transfer speeds, other capabilities of the \ac{ue}, information about the manufacturer of the wireless interface, and in some cases \ac{wps} field. 

\begin{figure}[tb]
    \centering
    \tikzset{every picture/.style={line width=0.75pt}} 

\begin{tikzpicture}[x=0.75pt,y=0.75pt,yscale=-1,xscale=1]

\draw    (196,472) .. controls (253.13,472.41) and (192.86,317.18) .. (253.16,312.13) ;
\draw [shift={(256,312)}, rotate = 179.63] [fill={rgb, 255:red, 0; green, 0; blue, 0 }  ][line width=0.08]  [draw opacity=0] (10.72,-5.15) -- (0,0) -- (10.72,5.15) -- (7.12,0) -- cycle    ;
\draw [line width=1.5]    (12,260) -- (448,260) ;
\draw    (196,500) .. controls (253.13,500.41) and (192.86,531.46) .. (253.16,532) ;
\draw [shift={(256,532)}, rotate = 179.63] [fill={rgb, 255:red, 0; green, 0; blue, 0 }  ][line width=0.08]  [draw opacity=0] (10.72,-5.15) -- (0,0) -- (10.72,5.15) -- (7.12,0) -- cycle    ;

\draw  [fill={rgb, 255:red, 255; green, 255; blue, 255 }  ,fill opacity=1 ]  (18,310) -- (198,310) -- (198,342) -- (18,342) -- cycle  ;
\draw (108,326) node   [align=left] {\begin{minipage}[lt]{119.68pt}\setlength\topsep{0pt}
\begin{center}
Duration
\end{center}

\end{minipage}};
\draw  [fill={rgb, 255:red, 255; green, 255; blue, 255 }  ,fill opacity=1 ]  (18,342) -- (198,342) -- (198,374) -- (18,374) -- cycle  ;
\draw (108,358) node   [align=left] {\begin{minipage}[lt]{119.68pt}\setlength\topsep{0pt}
\begin{center}
Destination Address
\end{center}

\end{minipage}};
\draw  [fill={rgb, 255:red, 255; green, 255; blue, 255 }  ,fill opacity=1 ]  (18,374) -- (198,374) -- (198,406) -- (18,406) -- cycle  ;
\draw (108,390) node   [align=left] {\begin{minipage}[lt]{119.68pt}\setlength\topsep{0pt}
\begin{center}
Source Address
\end{center}

\end{minipage}};
\draw  [fill={rgb, 255:red, 255; green, 255; blue, 255 }  ,fill opacity=1 ]  (18,406) -- (198,406) -- (198,438) -- (18,438) -- cycle  ;
\draw (108,422) node   [align=left] {\begin{minipage}[lt]{119.68pt}\setlength\topsep{0pt}
\begin{center}
BSS ID
\end{center}

\end{minipage}};
\draw  [fill={rgb, 255:red, 255; green, 255; blue, 255 }  ,fill opacity=1 ]  (18,278) -- (198,278) -- (198,310) -- (18,310) -- cycle  ;
\draw (108,294) node   [align=left] {\begin{minipage}[lt]{119.68pt}\setlength\topsep{0pt}
\begin{center}
Frame Control = 4
\end{center}

\end{minipage}};
\draw  [fill={rgb, 255:red, 255; green, 255; blue, 255 }  ,fill opacity=1 ]  (18,470) -- (198,470) -- (198,502) -- (18,502) -- cycle  ;
\draw (108,486) node   [align=left] {\begin{minipage}[lt]{119.68pt}\setlength\topsep{0pt}
\begin{center}
Information Element
\end{center}

\end{minipage}};
\draw  [fill={rgb, 255:red, 255; green, 255; blue, 255 }  ,fill opacity=1 ]  (254,310) -- (442,310) -- (442,342) -- (254,342) -- cycle  ;
\draw (348,326) node   [align=left] {\begin{minipage}[lt]{125.12pt}\setlength\topsep{0pt}
\begin{center}
SSID
\end{center}

\end{minipage}};
\draw  [fill={rgb, 255:red, 255; green, 255; blue, 255 }  ,fill opacity=1 ]  (254,342) -- (442,342) -- (442,374) -- (254,374) -- cycle  ;
\draw (348,358) node   [align=left] {\begin{minipage}[lt]{125.12pt}\setlength\topsep{0pt}
\begin{center}
Supported Rates
\end{center}

\end{minipage}};
\draw  [fill={rgb, 255:red, 255; green, 255; blue, 255 }  ,fill opacity=1 ]  (254,406) -- (442,406) -- (442,438) -- (254,438) -- cycle  ;
\draw (348,422) node   [align=left] {\begin{minipage}[lt]{125.12pt}\setlength\topsep{0pt}
\begin{center}
HT Capabilities
\end{center}

\end{minipage}};
\draw  [fill={rgb, 255:red, 255; green, 255; blue, 255 }  ,fill opacity=1 ]  (254,438) -- (442,438) -- (442,470) -- (254,470) -- cycle  ;
\draw (348,454) node   [align=left] {\begin{minipage}[lt]{125.12pt}\setlength\topsep{0pt}
\begin{center}
VHT Capabilities
\end{center}

\end{minipage}};
\draw  [fill={rgb, 255:red, 255; green, 255; blue, 255 }  ,fill opacity=1 ]  (254,470) -- (442,470) -- (442,502) -- (254,502) -- cycle  ;
\draw (348,486) node   [align=left] {\begin{minipage}[lt]{125.12pt}\setlength\topsep{0pt}
\begin{center}
Vendor Specific
\end{center}

\end{minipage}};
\draw  [fill={rgb, 255:red, 255; green, 255; blue, 255 }  ,fill opacity=1 ]  (254,374) -- (442,374) -- (442,406) -- (254,406) -- cycle  ;
\draw (348,390) node   [align=left] {\begin{minipage}[lt]{125.12pt}\setlength\topsep{0pt}
\begin{center}
Extended Supported Rates
\end{center}

\end{minipage}};
\draw  [fill={rgb, 255:red, 255; green, 255; blue, 255 }  ,fill opacity=1 ]  (18,438) -- (198,438) -- (198,470) -- (18,470) -- cycle  ;
\draw (108,454) node   [align=left] {\begin{minipage}[lt]{119.68pt}\setlength\topsep{0pt}
\begin{center}
Sequence Number
\end{center}

\end{minipage}};
\draw  [draw opacity=0]  (254,278) -- (442,278) -- (442,310) -- (254,310) -- cycle  ;
\draw (348,294) node   [align=left] {\begin{minipage}[lt]{125.12pt}\setlength\topsep{0pt}
\begin{center}
Optional
\end{center}

\end{minipage}};
\draw  [fill={rgb, 255:red, 255; green, 255; blue, 255 }  ,fill opacity=1 ]  (230,182) -- (442,182) -- (442,214) -- (230,214) -- cycle  ;
\draw (336,198) node   [align=left] {\begin{minipage}[lt]{141.44pt}\setlength\topsep{0pt}
\begin{center}
Channel
\end{center}

\end{minipage}};
\draw  [fill={rgb, 255:red, 255; green, 255; blue, 255 }  ,fill opacity=1 ]  (18,214) -- (230,214) -- (230,246) -- (18,246) -- cycle  ;
\draw (124,230) node   [align=left] {\begin{minipage}[lt]{141.44pt}\setlength\topsep{0pt}
\begin{center}
RSSI
\end{center}

\end{minipage}};
\draw  [fill={rgb, 255:red, 255; green, 255; blue, 255 }  ,fill opacity=1 ]  (18,182) -- (230,182) -- (230,214) -- (18,214) -- cycle  ;
\draw (124,198) node   [align=left] {\begin{minipage}[lt]{141.44pt}\setlength\topsep{0pt}
\begin{center}
Antenna
\end{center}

\end{minipage}};
\draw  [fill={rgb, 255:red, 255; green, 255; blue, 255 }  ,fill opacity=1 ]  (18,150) -- (230,150) -- (230,182) -- (18,182) -- cycle  ;
\draw (124,166) node   [align=left] {\begin{minipage}[lt]{141.44pt}\setlength\topsep{0pt}
\begin{center}
Time of Arrival
\end{center}

\end{minipage}};
\draw  [fill={rgb, 255:red, 255; green, 255; blue, 255 }  ,fill opacity=1 ]  (230,214) -- (442,214) -- (442,246) -- (230,246) -- cycle  ;
\draw (336,230) node   [align=left] {\begin{minipage}[lt]{141.44pt}\setlength\topsep{0pt}
\begin{center}
More Optional Information
\end{center}

\end{minipage}};
\draw  [draw opacity=0]  (230,150) -- (442,150) -- (442,182) -- (230,182) -- cycle  ;
\draw (336,166) node   [align=left] {\begin{minipage}[lt]{141.44pt}\setlength\topsep{0pt}
\begin{center}
Radio Interface Information
\end{center}

\end{minipage}};
\draw  [fill={rgb, 255:red, 255; green, 255; blue, 255 }  ,fill opacity=1 ]  (254,502) -- (442,502) -- (442,534) -- (254,534) -- cycle  ;
\draw (348,518) node   [align=left] {\begin{minipage}[lt]{125.12pt}\setlength\topsep{0pt}
\begin{center}
More optional Information
\end{center}

\end{minipage}};

\end{tikzpicture}
 
    \caption{Probe request structure.}
    \label{fig:probe_structure}
\end{figure}
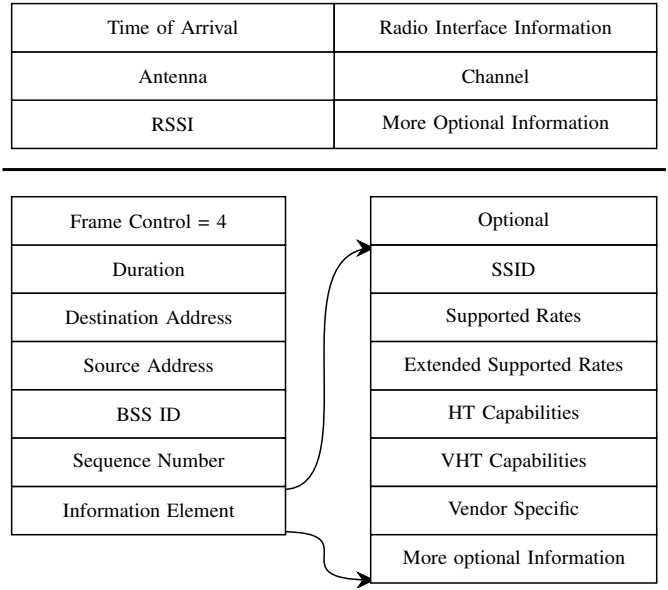

The probe request frame does not contain information about the transmission time, which means the \ac{toa} is taken from the \ac{rtc} timer of the receiver. Similar to the \ac{toa}, but dependent on the implementation of the capture device, the radio information can be extracted from the wireless interface. The most interesting information that can be gathered from the wireless interface is the channel on which the frame was received, to which antenna~in case the system has more than one, and the \ac{rssi} of the captured frame. The structure of the probe request, including the information gathered by the wireless interface of the receiver is visualized in Fig.~\ref{fig:probe_structure}.

\subsection{Existing Probe Request Datasets}

There are several publicly available datasets of probe requests. The datasets are described in the following list, including their advantages and disadvantages:

\begin{itemize}
    \item \textbf{Sapienza2013:} is the most commonly known probe request datasets~\cite{barbera2013crawdad, barbera2013signals}. It was published in 2013 which is before the implementation of \ac{mac} randomization algorithms in iOS, Android, or Windows. This means the dataset does not provide the same information as would be found in the real world. The anonymization in this dataset uses pseudonyms for \ac{ssid}s.
    \item \textbf{Glimps2015:} was captured at Glimps 2015, a~music festival that took place in Ghent, Belgium~\cite{hasselt-glimps2015-CRAWDAD}. There are several issues with this dataset. First, only one probe request is collected per unique \ac{mac} address, as well as timing issues due to the capture style. The anonymization techniques used in this dataset prove problematic for analysis as well. The dataset uses hashes for most of the information element fields, however, all of the sequence numbers are set to 0 and all \ac{ssid}s are hidden. This makes the analysis of this dataset quite difficult or straight-up impossible as the data of interest are hidden or missing.
    \item \textbf{Nile2021:} was captured during the night in a~shopping center and it is the only dataset~\cite{nile2021probes} not applying any anonymization which is perfect for analysis, however, the length of the data~capture is just 40 minutes. This is limiting to the analysis.
    \item \textbf{IPIN2021:} is our previous dataset,  published as supplementary materials for a~case study performed at IPIN 2021~\cite{bravenec2022zenodoIpin, bravenec2022your}. The dataset was collected over the course of only 4 days and our capture device lacked the capability of storing the radio information as well. The anonymization of this dataset was done using SHA512 overall privacy-sensitive fields, which does not prevent analysis while preserving user privacy.
\end{itemize}

In this work, we have decided to create a~new probe request dataset without the previous shortcomings. This dataset covers the time of 1 month and uses hashing for anonymization. This preserves user privacy and makes it impossible to match the data to real people while allowing full analysis on the same scale as non-anonymized data.

\subsection{Outline}

The paper is divided into several sections, in Section~\ref{section:dataset_creation} we discuss the technology used for the creation of the dataset. Following is the Section~\ref{section:data_description} describing the captured data. Some of the use case examples are presented in Section~\ref{section:examples}. The very important discussion about ethics is in Section~\ref{section:ethics}. And finally, the conclusions and brief look at our future work are in Section~\ref{section:conclusions}.

\begin{figure}[tb]
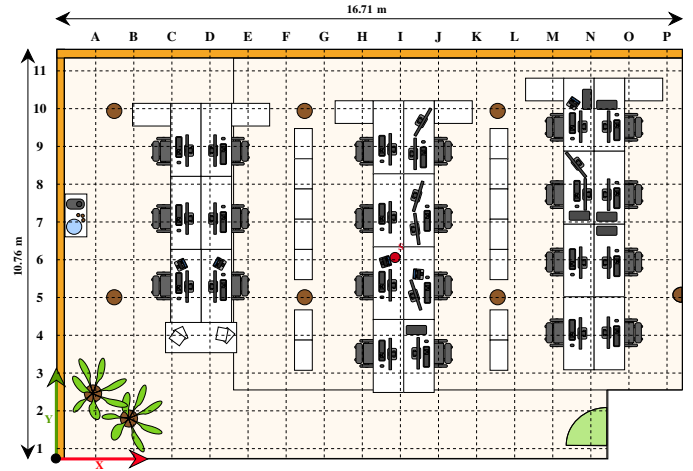

    \centering
    \includestandalone[width=\hsize]{images/init_map}
    \caption{Floor plan and location of sniffer in the office space of GEOTEC department at UJI, Spain.}
    \label{fig:office_space}
\end{figure}

\section{Dataset Creation}
\label{section:dataset_creation}

The dataset was collected over the time span of 1 month, specifically during the month of March 2023, in our office at University Jaume I, Spain. For collection, the ESP32-based sniffer from our previous works~\cite{bravenec2022your, bravenec2022exploration, fryza2023security} was used. The firmware for the ESP32 and all of its variants are publicly available from GitLab repository~\cite{bravenec2022gitlabSniffer}.

The sniffer captures raw packets and stores them in a~standardized packet capture file, compatible with network analysis tools like Wireshark, Python package scapy etc. The radio information is not part of the official 802.11 frames, as such we implemented saving of the Radiotap headers~\cite{radiotap} in conjunction with the 8o2.11 frames. This ensures compatibility with any packet analysis tools while providing us with additional information about the radio properties of the captured frame as well as preserving all captured information in the frame.

\section{Data~description}
\label{section:data_description}

As mentioned before, the dataset was collected in the office of the GEOTEC department of University Jaume I, Spain. The office has a~rectangular shape with dimensions of~\SI{16.71}{\meter} by~\SI{10.76}{\meter}. The office is of open-space style with a~typical occupancy of 14-20 people. The floor plan of the office is depicted in Fig.~\ref{fig:office_space}.

\begin{figure}[tb]
    \centering
    \includegraphics[width=\hsize]{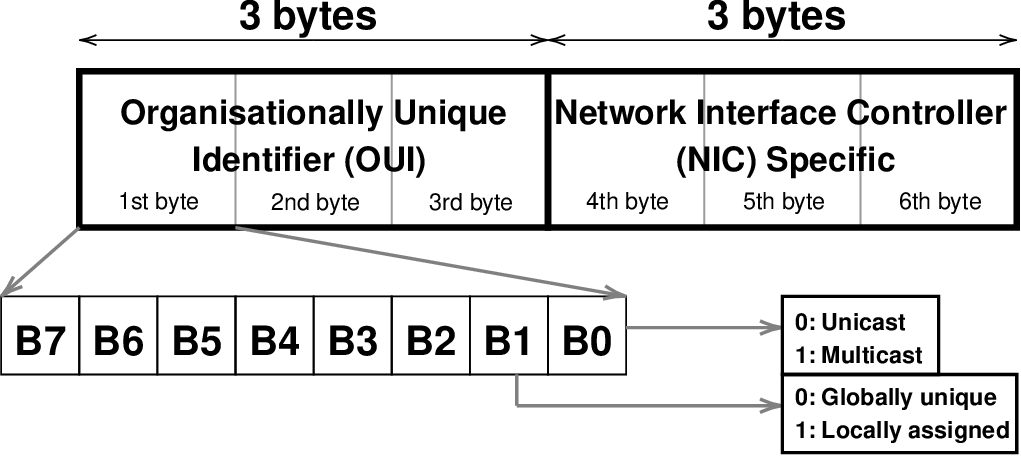}
    \caption{Structure of \ac{mac} address with the functional bits.}
    \label{fig:mac_address_structure}
\end{figure}

\begin{figure*}[tb]
    \centering
    \includegraphics[width=\hsize]{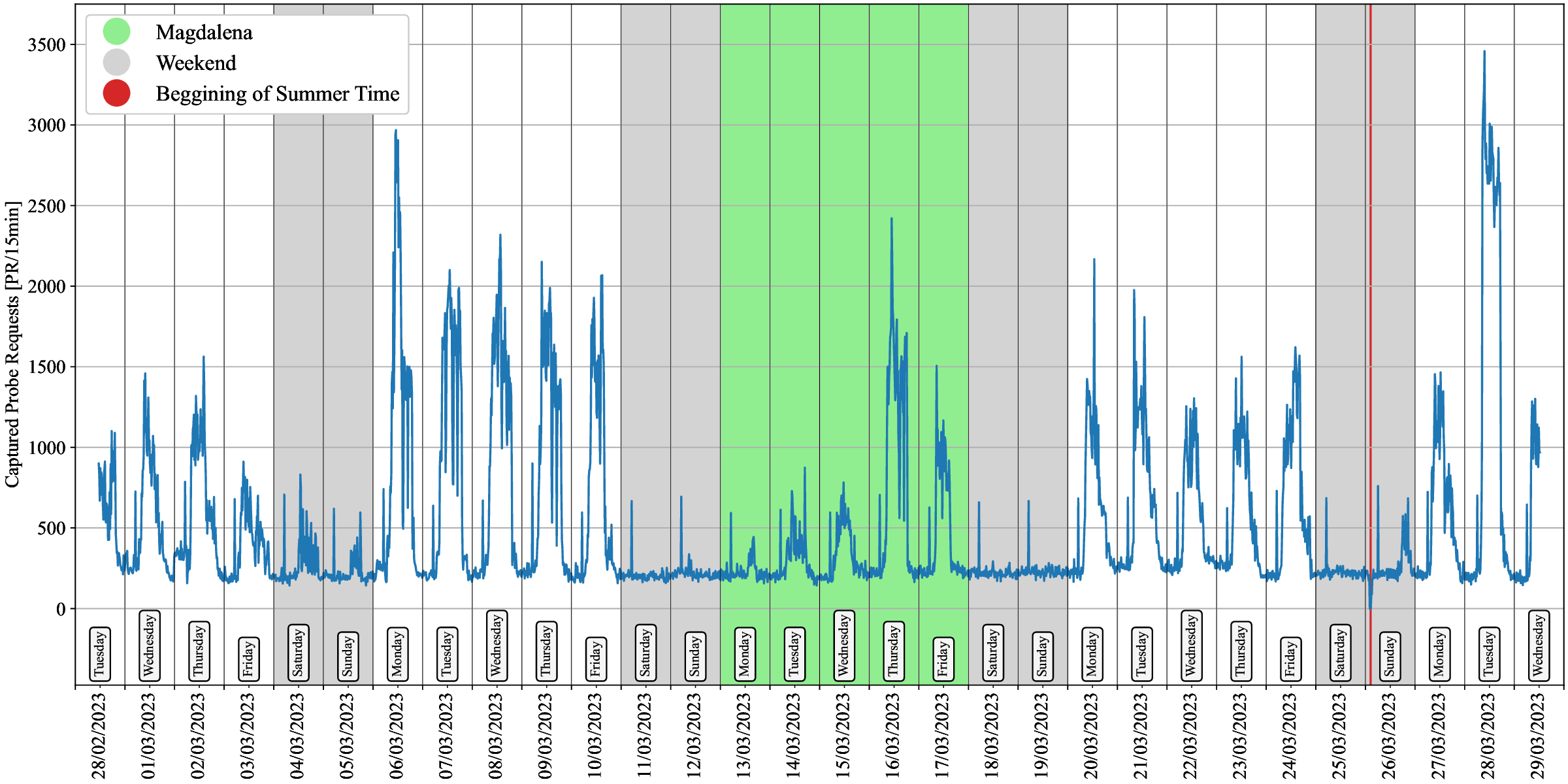}
    \caption{Density of captured Probe Requests over the course of capture (amount of probe requests grouped in 15-minute clusters).}
    \label{fig:probe_density}
\end{figure*}

\begin{figure}[tb]
    \centering
    \includegraphics[width=\hsize]{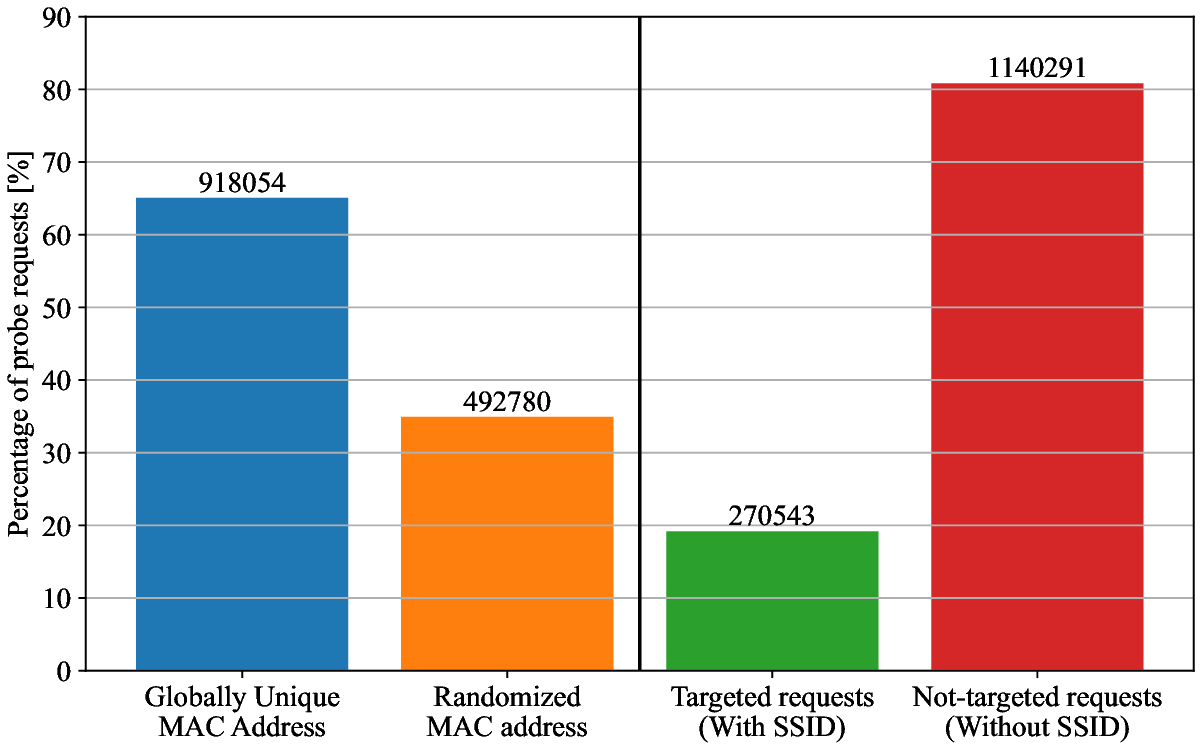}
    \caption{Split between randomized and globally unique \ac{mac} addresses in the captured probe requests and split between targeted and not targeted probe requests.}
    \label{fig:mac_split}
\end{figure}

The probe requests were collected during the month of March, to capture regular work weeks and also a~local holiday, Magdalena~2023, during which the university was mostly closed. Magdalena~2023 started on \formatdate{11}{03}{2023} and ended a~week later on \formatdate{19}{03}{2023}. The reduced occupancy of the office is clearly visible from Fig.~\ref{fig:probe_density}, where the holiday of Magdalena~are visualized in light green color.

There are some noticeable trends present in the dataset, that are visible in Fig.~\ref{fig:probe_density}. One of these trends is the noticeable constant transmission of probe requests. During the entire collection time, probe requests were being sent at all times, be it during the day, night, workday, or weekends. The explanation for the probe requests captured at night can be all-in-one computers using \wifi{} instead of a~wired connection, phones, or \ac{iot} devices used for experiments in the office that can be transmitting probe requests at all times. The second noticeable thing is a~peak in the transmitted probe requests happening every day around \formattime{7}{0}{0}. This is due to the scheduled reboot of a~\wifi{} access point present in the office, and devices searching for a~network to connect to after being disconnected from it. Another noticeable thing is a~short time period at night of the \formatdate{26}{03}{2023} with no captured probe requests. That is due to the switch to the Summer Time when the time changed from \formattime{2}{0}{0} to \formattime{3}{0}{0}.

We can also observe that some of our colleagues went to the office during some of the days of Magdalena~and during the first weekend since the beginning of the monitoring. This explains the lower, but still detectable, increase in captured probe requests.

\subsection{MAC addresses and SSIDs}

Another important factor in the analysis of probe requests is \ac{mac} addresses of devices. In case of devices not using randomization of \ac{mac} addresses, such \ac{ue} is very easy to track. Randomized \ac{mac} address is also very easily identifiable, due to the 2nd least significant bit \textbf{B1} in the first byte of the \ac{mac} address. When this bit is set, the \ac{mac} address was randomized by the network controller of the \ac{ue}. The least significant bit of the first byte \textbf{B0} distinguishes  individual devices and device groups. The structure of the \ac{mac} address is presented in Fig.~\ref{fig:mac_address_structure}. Considering this limitation of the 2 least significant bits, the 2nd digit of locally assigned \ac{mac} address in hexadecimal format has only four options: \num{2} (0010), \num{6} (0110), a~(1010) or E (1110). From the captured probe requests, about \SI{35}{\percent} used randomized \ac{mac} addresses, which is presented in Fig.~\ref{fig:mac_split}.

The second very important parameter is the \ac{pnl}, which is the list of network \ac{ssid}s the \ac{ue} often connects to. This can be leaked by the device sending probe requests targeted for specific networks. In this dataset, \SI{19}{\percent} of the captured probe requests contained an \ac{ssid}, which is presented in Fig.~\ref{fig:mac_split}. Even though this seems like a~fairly low number, the dataset contains 2030 unique \ac{ssid}s.

\subsection{Information Elements}

This optional section of probe requests is very useful for fingerprinting, as it presents us with a~set of capabilities and supported functions that are not changing over time for a~single \ac{ue}. The information ranges from supported data~rates, capabilities related to certain \wifi{} standards (High Throughput (HT) capabilities -802.11n, Very High Throughput (VHT) capabilities - 802.11ac, and High Efficiency (HE) capabilities - 802.11ax), vendor-specific elements (some devices had several of them) to \ac{wps} fields which can contain many user identifying information ranging from the device manufacturer, all the way to device name (and since many devices are named by users - \textit{Julia's iPhone}, this can leak the name of the user). \ac{wps} fields also come with \ac{uuide},  also a~unique identifier that can compromise the anonymity while using randomized \ac{mac} addresses as it does not change over time. The occurrences of each field are shown in Table~\ref{tab:probe_fields}.


\begin{table}[bt]
    \centering
    \caption{Probe request fields used to create device fingerprint and frequency of occurrence in data~collected in our lab}
    \label{tab:probe_fields}
    \begin{tabular*}{\columnwidth}{llS[table-format=7]S[table-format=3.2]}
    \toprule     
        \multicolumn{2}{c}{Information Element}            & {Included in Probes}      & [\SI{}{\percent}] \\ \midrule
        \multicolumn{2}{l}{Supported rates}                & 1410832                    & 100.00     \\
        \multicolumn{2}{l}{Extended Supported rates}       & 1405288                    &  99.61     \\
        \multicolumn{2}{l}{HT Capabilities}                & 1093575                    &  77.51     \\
        \multicolumn{2}{l}{HE Capabilities}                &  394347                    &  27.95     \\
        \multicolumn{2}{l}{VHT Capabilities}               &  262365                    &  18.60     \\
        \multicolumn{2}{l}{Extended Capabilities}          & 1182014                    &  83.78     \\
        \multicolumn{2}{l}{Vendor Specific elements}       &  315815                    &  22.39     \\
        & 1 Vendor Specific element                        &  135973                    &   9.64     \\
        & 2 Vendor Specific elements                       &  110071                    &   7.80     \\
        & 3 Vendor Specific elements                       &    9607                    &   0.68     \\
        & 4 Vendor Specific elements                       &     257                    &   0.02     \\
        & 5+ Vendor Specific elements                      &     146                    &   0.01     \\
        \multicolumn{2}{l}{WPS - UUID-E}                   &   16596                    &   1.18     \\
        \multicolumn{2}{l}{WEP Protected}                  &       2                    &   0.00     \\ \midrule
        \multicolumn{2}{l}{Total Collected Probe Requests} & \multicolumn{2}{S[table-format=12.0]}{1410834} \\ \bottomrule
    \end{tabular*}
\end{table}

\subsection{Radio Information}

The ESP32 sniffer was configured to gather probe requests in all channel scan. Due to the lack of support of ESP32 for the \SI{5}{\giga\hertz} band of \wifi{}, only probe requests transmitted in the \SI{2.4}{\giga\hertz} band were captured.

\begin{figure}[!tb]
    \centering
    \includegraphics[width=\hsize]{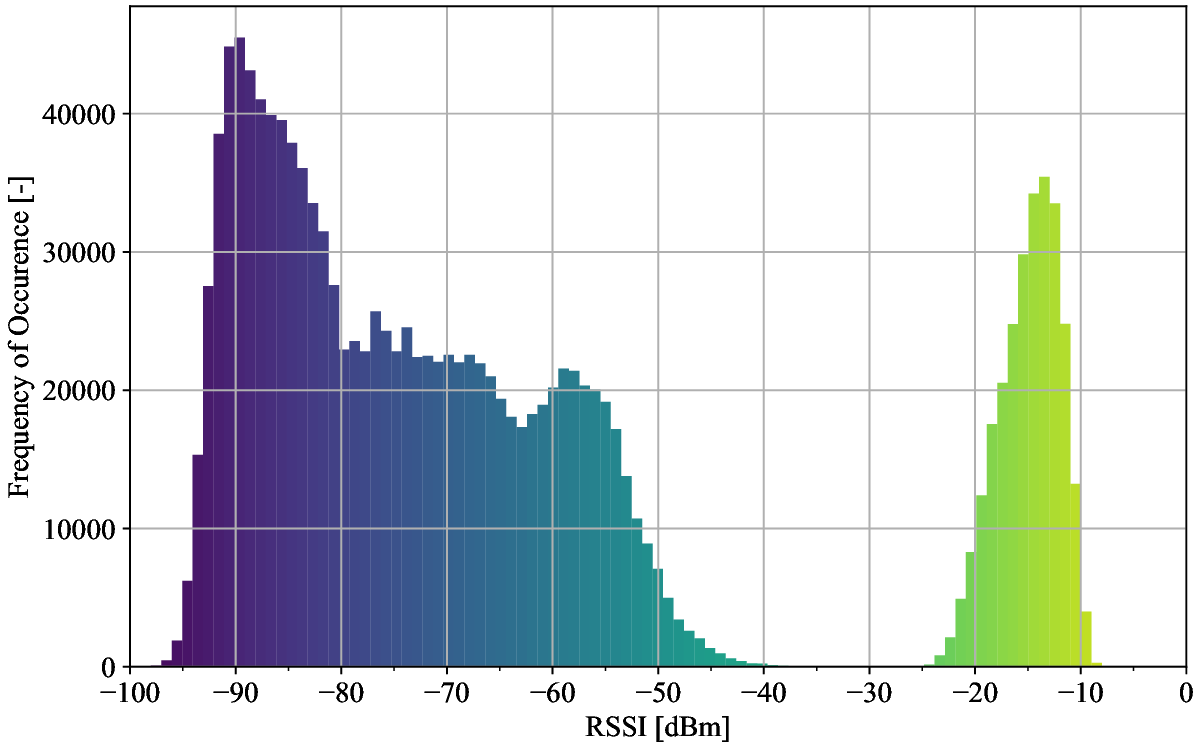}
    \caption{Frequency of occurrence of \ac{rssi} in the captured probe requests.}
    \label{fig:rssi_histogram}
    \vspace{20pt}
    \includegraphics[width=\hsize]{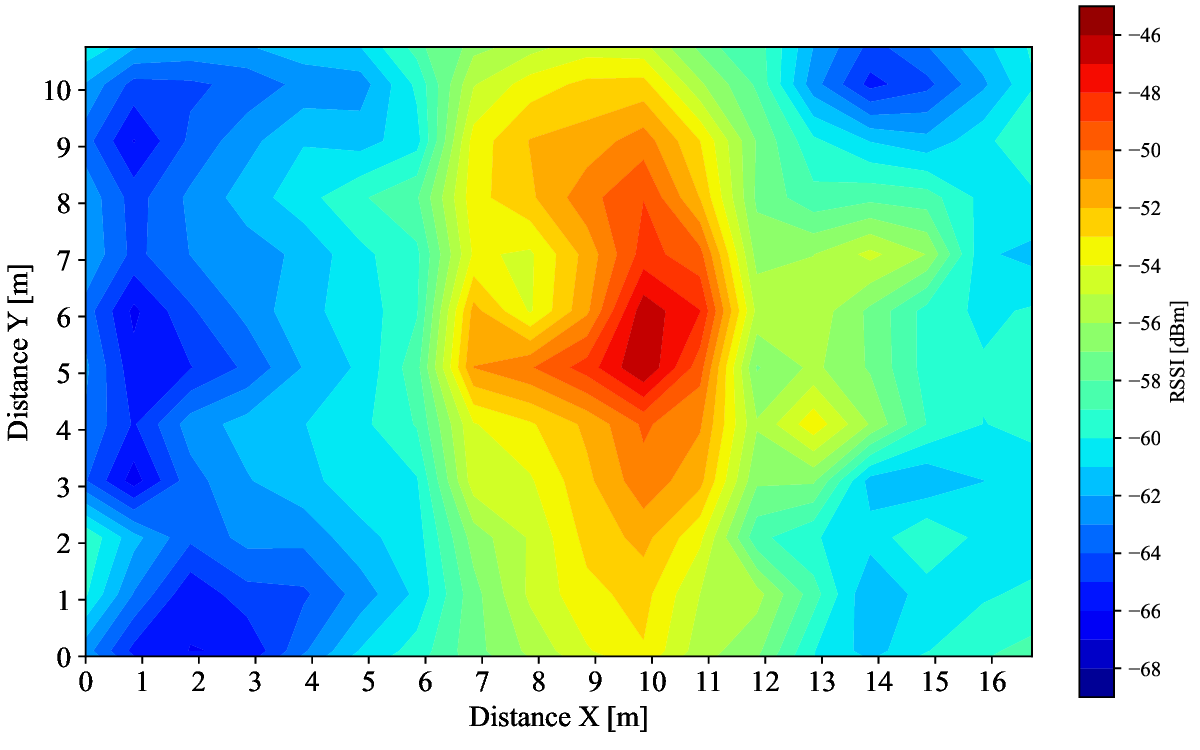}
    \caption{\acl{rem} of the \ac{rssi} in different locations of the office mapped with ESP32 micro controllers.}
    \label{fig:rssi_rem}
    \vspace{20pt}
    \includegraphics[width=\hsize]{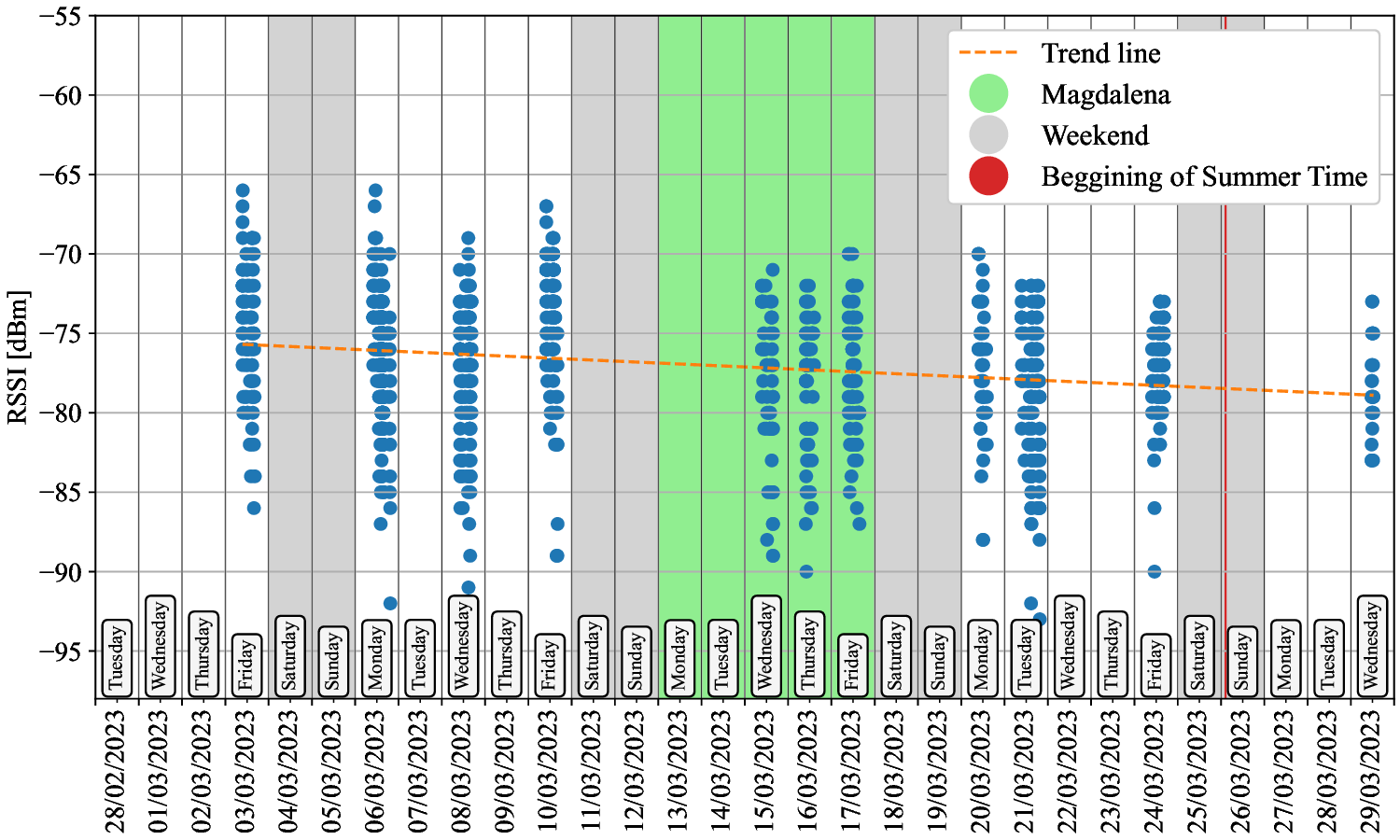}
    \caption{Example of \ac{rssi} long-term stability evaluation for one \ac{mac} address from the dataset (\textit{f4:7b:09:1f:5b:72}) including trend line.}
    \label{fig:rssi_stability}
\end{figure}

The distribution of \ac{rssi} values are presented using a~histogram in Fig.~\ref{fig:rssi_histogram}. From the histogram, it is clearly visible that most frames were captured with \ac{rssi} in the range of \SIrange{-100}{-40}{dBm}.  Surprisingly, a~lot of probe requests had \ac{rssi} higher than \SI{-30}{dBm}: all of these probes were transmitted by a~single device that had the wireless interface configured to use higher transmit power than usual. Another point of interest is the high amount of probes captured with the \ac{rssi} around \SI{-90}{dBm}, which are most likely devices from neighboring offices. We have also mapped the radio signal propagation throughout our office for another study. The \ac{rem} of the office mapped using another ESP32 microcontroller is in Fig.~\ref{fig:rssi_rem}. The capture \ac{rem} can be used as a~rough distance filter, as it can be used for the selection of \ac{rssi} threshold.

\section{Usage Examples}
\label{section:examples}

There are several ways to use the dataset. In this section, we provide just a~few examples of possible use cases.

\subsection{Wi-Fi Signal Stability Evaluation}

By capturing the \ac{rssi} values for each captured probe request, it is possible to analyze the signal from the received signal strength point of view as well\added{, assuming people spend most of the time at their desks}. In the example we propose we measure the signal stability over time. As an example, we have randomly selected \ac{mac} address of a~single device that appears on several days (\ac{mac} address of the selected device in the anonymized dataset is \textit{f4:7b:09:1f:5b:72}) and visualized the \ac{rssi} of all probe requests over time. The changes over time are visible, and to highlight the change we also included a~trend line in a~Fig.~\ref{fig:rssi_stability} representing the decrease in \ac{rssi} throughout the capture time.

\subsection{Presence Detection \& Room Occupancy Estimation}

The network traffic can be also used for presence detection. This can be easily seen in Fig.~\ref{fig:probe_density}, from where we can see when the activity in the proximity of the sniffer increased. We can also see the drops on weekends and holidays. By also using the \ac{rem} from Fig.~\ref{fig:rssi_rem}, it is possible to roughly estimate the distance the \ac{ue} is from the sniffer. \added{The \ac{rem} was created by collecting \ac{rssi} of probe requests throughout the office in \SI{1}{\meter} grid.} This information can be utilized for presence detection, or even room occupancy detection. On a~smaller scale, we have estimated the occupancy of rooms on the university campus from captured probe requests~\cite{fryza2023security}.

Unfortunately, the capture of the ground truth of room occupancy in our office is nearly impossible. With up to 30 people present in the office, moving in and out at all times during work hours, the collection of ground truth was not possible with our current capabilities. However, the occupancy can be classified like: \textit{empty, low, medium, high}, or on a~scale from \textit{low} to \textit{high} which was used previously in occupation estimation work by \textcite{ciftler2017occupancy}.

\subsection{User Privacy Exploitation}

Since during the anonymization of the dataset we did not remove any data, but used pseudonyms using hashing for sensitive fields, the dataset allows for analysis of privacy leakage using an up to date data. The \ac{toa} of packets was not modified in any way either, which allows for studies into temporal analysis similar to the work done by \textcite{matte2016defeating}.

It can also be used for the analysis of randomized \ac{mac} address recurrences or identification of users despite the randomization. The dataset may be used to reveal present vulnerabilities of the probe request mechanism and creation of a~new type of analysis.

\section{Ethics and Sensitive Information}
\label{section:ethics}

The output packet capture files created by the ESP32-based probe request sniffer are in the same style as when the packets are captured by network analysis tools like Wireshark. The captured packets are the same as those transmitted by the nearby devices, including any sensitive information that they might contain. This information might contain globally unique \ac{mac} addresses, randomized \ac{mac} addresses, \ac{ssid}s of networks saved in the devices' preferred network list, manufacturer of the device or the wireless network interface, or in some cases even device names.

Following the data capture, we have run an anonymization script to hide any user information. For this purpose, the SHA512 hashing algorithm was used. By employing the hashing algorithm, we replace the original fields with hashes, which preserves the ability to analyze the data~in the same way as data that have not been anonymized. The main reason for anonymization through hashing is to allow the analysis of the probe requests to unveil the vulnerabilities in the management frames of the current implementation of the 802.11 protocol. To also preserve the information about randomized and globally unique \ac{mac} addresses, only the last \SI{3}{\byte} are hashed. This preserves the least significant bits of the 1st Byte of the \ac{mac} address, as well as the Organization Unique Identifier~\cite{ieeeOui}. Thanks to the use of hashing, creating a~connection between the real identities of people and the captured probe requests is not possible. The anonymized version of the dataset is publicly available from Zenodo repository~\cite{bravenec2023zenodoIpin}.

\section{Conclusions \& Future Work}
\label{section:conclusions}

In this paper, we have explored existing datasets of \wifi{} probe requests and presented a~new one. As \wifi{} enabled devices evolve every year, our new dataset provides updated data. Apart from the data~itself we also provide firmware for the ESP32 microcontroller, so anyone can collect their own dataset. We present the data-gathering methodology and describe important information about the features of the dataset. Additionally, we have taken into account possible use cases and presented possible outcomes, like temporal \ac{rssi} stability, presence detection, room occupancy estimation, and privacy breaches.

In future works, we will dive deeper into the \wifi{} in indoor positioning applications. We want especially to focus on the exploration of privacy issues and passive user tracking possibilities using the IEEE 802.11 communication protocol.

\balance
\renewcommand*{\bibfont}{\footnotesize}
\renewcommand*{\UrlFont}{\rmfamily}
\printbibliography
\end{document}